\documentclass[aps,prl,superscriptaddress,twocolumn,]{revtex4-2}
\pdfoutput=1
\usepackage{amssymb}
\usepackage{graphicx}
\usepackage{amsmath}
\usepackage[colorlinks=true,linkcolor=blue,citecolor=blue,urlcolor=blue]{hyperref}
\usepackage{tikz} 

\begin{document}

\title{Space-Time Duality between Quantum Chaos and Non-Hermitian Boundary Effect}
\author{Tian-Gang Zhou}
\thanks{They contribute equally to this work. }
\affiliation{Institute for Advanced Study, Tsinghua University, Beijing,100084, China}

\author{Yi-Neng Zhou}
\thanks{They contribute equally to this work. }
\affiliation{Institute for Advanced Study, Tsinghua University, Beijing,100084, China}

\author{Pengfei Zhang}
\thanks{PengfeiZhang.physics@gmail.com}
\affiliation{Institute for Quantum Information and Matter and Walter Burke Institute for Theoretical Physics, California Institute of Technology, Pasadena, California 91125, USA}

\author{Hui Zhai}
\thanks{hzhai@mail.tsinghua.edu.cn}
\affiliation{Institute for Advanced Study, Tsinghua University, Beijing,100084, China}
\date{\today}

\begin{abstract}

Quantum chaos in hermitian systems concerns the sensitivity of long-time dynamical evolution to initial conditions. The skin effect discovered recently in non-hermitian systems reveals the sensitivity to the spatial boundary condition even deeply in bulk. In this letter, we show that these two seemingly different phenomena can be unified through space-time duality. The intuition is that the space-time duality maps unitary dynamics to non-unitary dynamics and exchanges the temporal direction and spatial direction. Therefore, the space-time duality can establish the connection between the sensitivity to the initial condition in the temporal direction and the sensitivity to the boundary condition in the spatial direction. Here we demonstrate this connection by studying the space-time duality of the out-of-time-ordered commutator in a concrete chaotic hermitian model. We show that the out-of-time-ordered commutator is mapped to a special two-point correlator in a non-hermitian system in the dual picture. For comparison, we show that this sensitivity disappears when the non-hermiticity is removed in the dual picture.

\end{abstract}

\maketitle

{\color{blue}\emph{Introduction.}} Chaos describes the phenomenon that the future is highly sensitive to any small perturbation at present, and this sensitivity can be more significant for a longer evolution time. During the past years, chaos in quantum systems has been extensively studied in terms of the out-of-time-ordered commutator (OTOC), which shows that chaotic behavior is a general property in most quantum many-body systems \cite{larkin1969quasiclassical,almheiriApologiaFirewalls2013,shenkerBlackHolesButterfly2014,robertsDiagnosingChaosUsing2015,robertsLocalizedShocks2015,shenkerStringyEffectsScrambling2015,hosurChaosQuantumChannels2016a,maldacenaBoundChaos2016,stanfordManybodyChaosWeak2016a,hosurChaosQuantumChannels2016,chenTunableQuantumChaos2017a,chenCompetitionChaoticNonChaotic2017,songStronglyCorrelatedMetal2017a,maldacenaConformalSymmetryIts2016,maldacenaRemarksSachdevYeKitaevModel2016a,zhuMeasurementManybodyChaos2016,guLocalCriticalityDiffusion2016,patelQuantumButterflyEffect2017,vonkeyserlingkOperatorHydrodynamicsOTOCs2018a,xuLocalityQuantumFluctuations2019a,zhangQuantumChaosUnitary2019a,guRelationMagnitudeExponent2019,zhangObstacleSubAdSHolography2021,guoTransportChaosLattice2019,lewis-swanUnifyingScramblingThermalization2019a,miInformationScramblingComputationally2021}. As a separate development, the non-hermitian skin effect has been discovered recently as a generic feature in non-hermitian systems both theoretically \cite{yaoEdgeStatesTopological2018,kunstBiorthogonalBulkBoundaryCorrespondence2018,martinezalvarezNonHermitianRobustEdge2018,leeAnatomySkinModes2019} and experimentally \cite{garttnerMeasuringOutoftimeorderCorrelations2017,ghatakObservationNonHermitianTopology2020,helbigGeneralizedBulkBoundary2020,xiaoNonHermitianBulkBoundary2020}. The non-hermitian skin effect states that the eigenstates of a non-Hermitian Hamiltonian can be highly sensitive to the spatial boundary condition. Unlike hermitian systems, this sensitively holds even deeply in the bulk and far from the boundary.    

Despite that both effects concern the sensitivity to perturbations, they look pretty different at first glance. First, quantum chaos is mostly discussed in the hermitian system, and the skin effect is unique to the non-hermitian system. Secondly, quantum chaos concerns the sensitivity on the temporal domain, and the non-hermitian skin effect concerns the sensitivity on the spatial domain. Therefore, no previous discussion has brought out the connection between these two effects, not to speak of the possible equivalence between them. 

In this letter, we show that these two effects can be unified under the space-time duality of the quantum circuit. As we will review below, the space-time duality of the quantum circuit maps unitary dynamics to non-unitary dynamics and simultaneously exchanges the role of spatial direction and time direction. Therefore, it is very intuitive to understand that the space-time duality can bridge the gap between these two phenomena. Here we demonstrate such intuition with a concrete example. 

\begin{figure}
 	\centering
 	\includegraphics[width=0.9\linewidth]{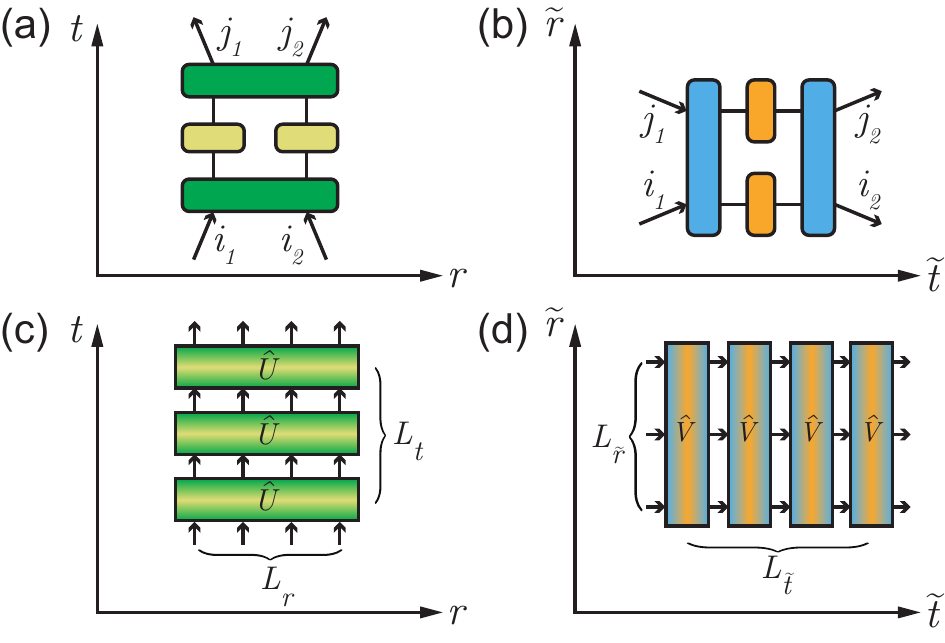}
 	\caption{Schematic of space-time duality of quantum circuit. (a, b): Space-time duality between two-qubit quantum circuits $\hat{u}$ (left) and $\hat{v}$ (right) with $v_{i_1,j_1}^{i_2,j_2}=u_{i_1,i_2}^{j_1,j_2}$. Green and blue box respectively represent two-qubit gate $e^{iJ_z\hat{\sigma}^z_1\hat{\sigma}^z_2}$ and $e^{i\tilde{J}_z\hat{\sigma}^z_1\hat{\sigma}^z_2}$. Yellow and orange box respectively represent single-qubit gate $e^{iJ_x\hat{\sigma}_x}$ and $e^{i\tilde{J}_x\hat{\sigma}_x}$, with the relation between $J_z$ and $\tilde{J}_z$, and relation between $J_x$ and $\tilde{J}_x$ given by Eq.~\eqref{relation}. (c, d): Space-time duality between two general operators $\hat{U}$ and $\hat{V}$, with $L_{\tilde{r}}=L_t$ and $L_{\tilde{t}}=L_r$.	}
 	\label{fig:Duality}
 \end{figure}
 
{\color{blue}\emph{Review of Space-Time Duality.}} Before proceeding, let us first briefly review the space-time duality of quantum circuit \cite{akilaParticletimeDualityKicked2016,bertiniExactSpectralForm2018,bertiniEntanglementSpreadingMinimal2019,bertiniExactCorrelationFunctions2019,gopalakrishnanUnitaryCircuitsFinite2019,claeysMaximumVelocityQuantum2020,piroliExactDynamicsDualunitary2020,kosCorrelationsPerturbedDualUnitary2021,fritzschEigenstateThermalizationDualunitary2021,ippolitiPostselectionFreeEntanglementDynamics2021,luEntanglementTransitionsSpacetime2021}. In the simplest case, let us consider a two-qubit gate $\hat{u}$ operating on a two-qubit state $|i_1\rangle\otimes|i_2\rangle$, and $\hat{u}|i_1\rangle\otimes|i_2\rangle=u_{i_1,i_2}^{j_1,j_2}|j_1\rangle\otimes|j_2\rangle$. Note that here we have fixed a given set of basis. In this case two qubits represent spatial direction and the incoming and outgoing represents the temporal direction. By exchanging the role of spatial and temporal directions, we define another operator $\hat{v}$, which acts as $\hat{v}|i_1\rangle\otimes|j_1\rangle=u_{i_1,i_2}^{j_1,j_2}|i_2\rangle\otimes|j_2\rangle$. That is to say, $\hat{v}$ called as the space-time duality circuit of $\hat{u}$ if $v_{i_1,j_1}^{i_2,j_2}=u_{i_1,i_2}^{j_1,j_2}$. One example is shown in Fig.~1, if we choose 
\begin{equation}
\hat{u}=e^{iJ_z\hat{\sigma}^z_1\hat{\sigma}^z_2}e^{iJ_x(\hat{\sigma}^x_1+\hat{\sigma}^x_2)}e^{iJ_z\hat{\sigma}^z_1\hat{\sigma}^z_2}, 
\end{equation}
and we fix the basis as the eigenbasis of $\hat{\sigma}^z$, it can be shown that the corresponding $\hat{v}$ has the same form as $\hat{u}$
\begin{equation}
\hat{v}=e^{i\tilde{J}_z\hat{\sigma}^z_1\hat{\sigma}^z_2}e^{i\tilde{J}_x(\hat{\sigma}^x_1+\hat{\sigma}^x_2)}e^{i\tilde{J}_z\hat{\sigma}^z_1\hat{\sigma}^z_2},
\end{equation}
and the parameters $\tilde{J}_x$ and $\tilde{J}_z$ are given by \cite{supple}
\begin{equation}
\tilde{J}_{x} = \arctan(-i e^{-2i J_z}),\   \  \tilde{J}_{z} = -\frac{\pi}{4} + \frac{i}{2} \ln(\tan J_x). \label{relation}
\end{equation} 
When $J_x$ and $J_z$ are both real and $\hat{u}$ is unitary, $\tilde{J}_x$ and $\tilde{J}_z$ are in general complex numbers, which means that $\hat{v}$ is a non-unitary evolution. 

More generally, as shown in Fig.~1(c) and (d), let us consider a unitary operator $\hat{U}$ repeatedly acting $L_t$ steps on this system with $L_r$ qubits, we can introduce a circuit $\hat{V}$ as space-time dual of $\hat{U}$, which repeatedly acts $L_{\tilde{t}}$ steps on a system with $L_{\tilde{r}}$ qubits. Here $L_{\tilde{r}}=L_t$ and $L_{\tilde{t}}=L_r$. For example, if we choose $\hat{U}$ as 
\begin{equation}
\hat{U}= e^{i \sum\limits_{r=1}^{L_r} J_{x} \hat{\sigma}^x_r} e^{i \sum\limits_{r=1}^{L_r} \left(J_{z} \hat{\sigma}^z_r \hat{\sigma}^z_{r+1} + h \hat{\sigma}^z_r\right)}, \label{bigU}
\end{equation}  
the corresponding $\hat{V}$ is given by (up to a constant) \cite{supple}
\begin{equation}
\hat{V}= e^{i \sum\limits_{r=1}^{L_{\tilde r}} \tilde{J}_{x} \hat{\sigma}^x_r} e^{i \sum\limits_{r=1}^{L_{\tilde r}} \left(\tilde{J}_{z} \hat{\sigma}^z_r \hat{\sigma}^z_{r+1} + h \hat{\sigma}^z_r\right)}, \label{bigV}
\end{equation}  
with the same relation between $\tilde{J}_x$, $\tilde{J}_z$ and $J_x$, $J_z$ as given by Eq.~\ref{relation}. In general, $\hat{V}$ is a non-unitary circuit when $\hat{U}$ is unitary.

\begin{figure}
 	\centering
 	\includegraphics[width=1.0\linewidth]{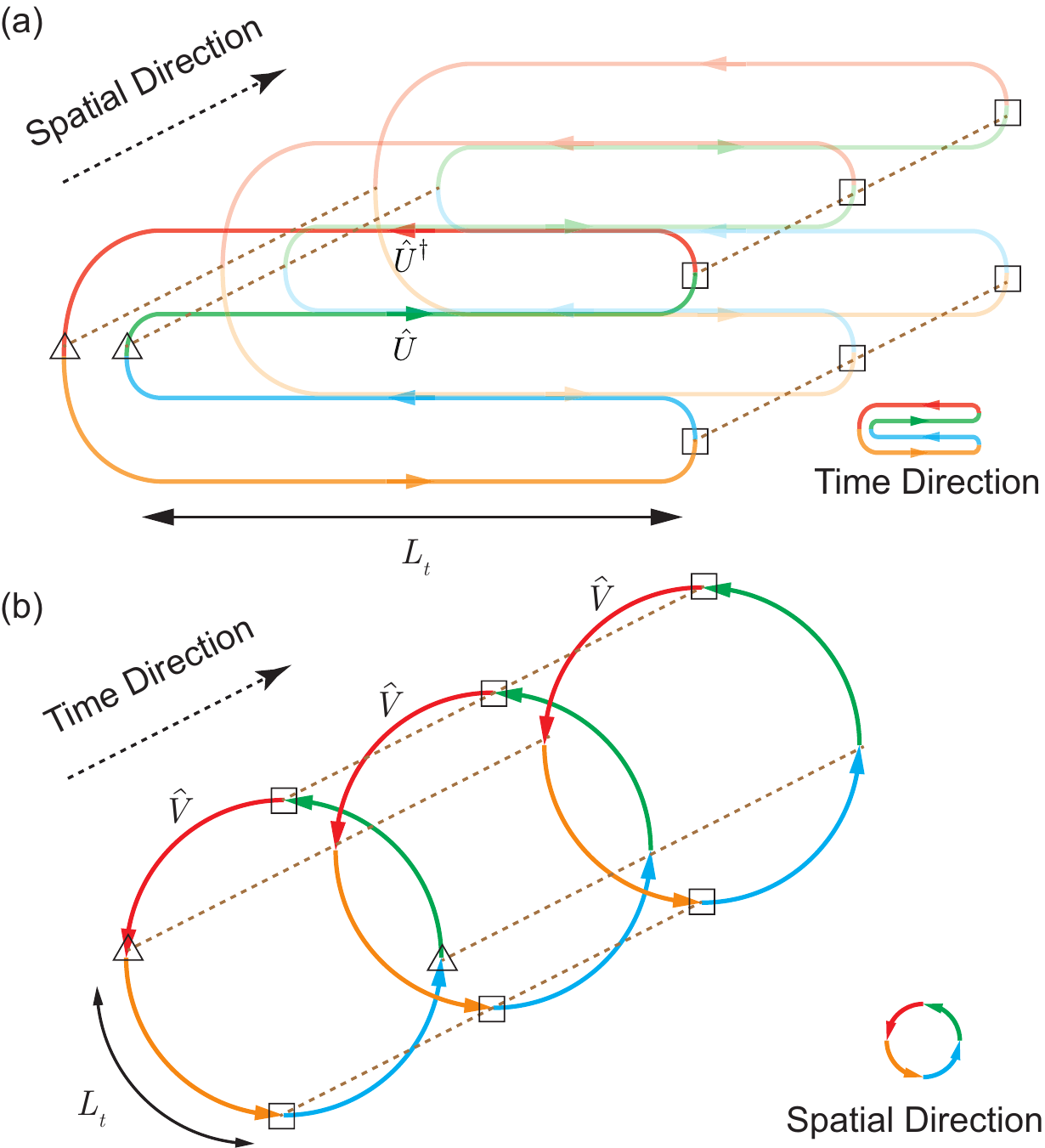}
 	\caption{(a) Correlation function $\mathcal{F}(\phi)$ defined in Eq.~\eqref{Fphi}. The solid line represents the temporal direction. Different branches of forward and backward evolutions in the double Keldysh contour are distinguished by different colors, and each part has a length $L_t$ (b) Space-time dual of $\mathcal{F}(\phi)$. The solid line now represents the spatial direction with $L_{\tilde{r}}=4L_t$. Squares and triangles are respectively label $e^{\pm i\phi\hat{W}}$ and $\hat{O}$ acting on the double Keldysh contour in (a) or their dual circuit acting on different spatial points on the spatial contour in (b). The distance between triangle and square is $L_t$ in both cases. }
 	\label{fig:OTOC}
 \end{figure}

{\color{blue}\emph{Space-Time Duality of OTOC.}} Here we first consider a function $\mathcal{F}(\phi)$ defined as
\begin{equation} \label{Fphi_origin}
\mathcal{F}(\phi)=\frac{1}{2^{L_r}}\text{Tr}_{L_r}\left[\hat{O}(L_t)e^{i\phi\hat{W}}\hat{O}^{\dag}(L_t)e^{-i\phi\hat{W}}\right]
\end{equation}
Here we choose $\hat{W}=\sum_r \hat w_r$ as an operator that uniformly acts on all spatial sites, where $\hat{w}_r$ denotes an operator $\hat{w}$ acting on site-$r$, and $\hat{O}$ to be a spatially local operator. The reasons we consider this correlation function are multifolds. First, this quantity is directly related to the OTOC. It can be shown that
\begin{equation}
\frac{\partial^2 \mathcal{F}(\phi)}{\partial \phi^2}\Big|_{\phi=0}=-\frac{1}{2^{L_r}}\text{Tr}_{L_r}\left[|[\hat{O}(L_t),\hat{W}]|^2\right], \label{derivative}
\end{equation}
and the r.h.s. of Eq.~\eqref{derivative} is the OTOC. Thus, for quantum chaos, the OTOC is larger for larger $L_t$, which means that $\mathcal{F}(\phi)$ should sensitively depend on $\phi$ even for larger $L_t$. Secondly, this quantity is closely related to the multiple quantum coherence that can be directly measured in NMR and trapped ion systems \cite{alvarezQuantumSimulationsLocalization2013,garttnerMeasuringOutoftimeorderCorrelations2017}. Thirdly, this quantity possesses a clear physical interpretation after performing space-time duality on the basis diagonal in $\hat{w_r}$, as we will discuss from the following three aspects. 

\begin{figure*}
	\centering
	\includegraphics[width=1.0\linewidth]{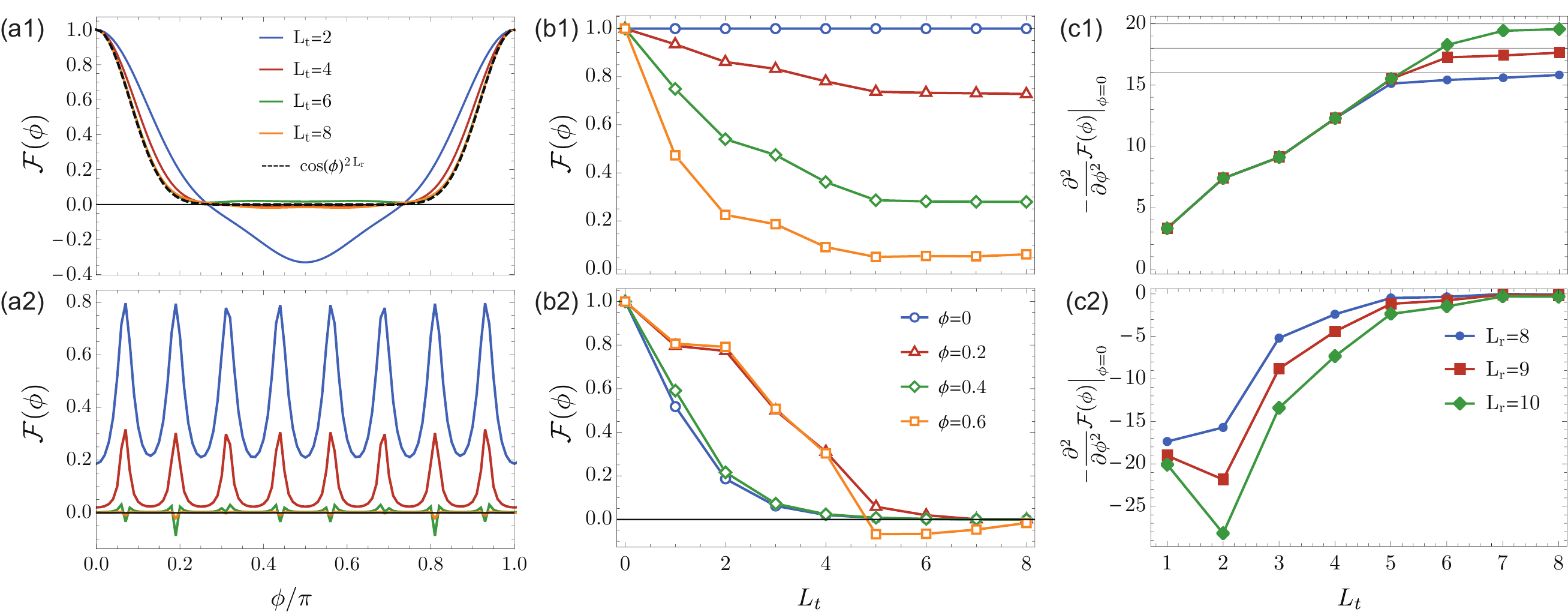}
	\caption{(a1,a2) $\mathcal{F}(\phi)$ as a function of $\phi$ for different $L_t$; (b1,b2) $\mathcal{F}(\phi)$ as a function of $L_t$ for different $\phi$; (c1,c2) $\partial^2\mathcal{F}(\phi)/\partial\phi^2|_{\phi=0}$ as a function of $L_t$ for different $L_r$. $L_r=8$ for (a1,a2) and (b1,b2). The top row plots the function given by Eq.~\eqref{Fphi}, or equivalently Eq.~\eqref{Fphid}, with $\hat{O}=\hat{\sigma}^z_1$ and $\hat{W}=\sum_{r=1}^{L_r}\hat{\sigma}^z_r$. We choose parameters $\{ J_{x}, J_{z}, h \}=\{ 1, 1, 0.5 \}$. The low row plots a modified Eq.~\eqref{Fphid} which eliminates the non-hermiticity (see text for details).
	}
	\label{fig:result}
\end{figure*}

i) Length of the dual spatial contour: In Eq.~\eqref{Fphi_origin}, $\hat{O}(L_t)$ is given by $(\hat{U}^{\dag})^{L_t}\hat{O}(\hat{U})^{L_t}$, and explicitly, $\mathcal{F}(\phi)$ can be written as 
\begin{equation}
\frac{1}{2^{L_r}}\text{Tr}_{L_r}\left[(\hat{U}^{\dag})^{L_t}\hat{O}(\hat{U})^{L_t}e^{i\phi\hat{W}}(\hat{U}^{\dag})^{L_t}\hat{O}(\hat{U})^{L_t}e^{-i\phi\hat{W}}\right]. \label{Fphi}
\end{equation}
Unlike the unidirectional evolution discussed above, $\mathcal{F}(\phi)$ contains two forward evolutions $(\hat{U}^{\dag})^{L_t}$ and two backward evolutions $(\hat{U})^{L_t}$. In other words, it contains two Keldysh contours. They are marked by different colors in Fig.~\ref{fig:OTOC}(a). The length of each evolution is $L_t$. Therefore, after space-time duality, the length $L_{\tilde{r}}$ of spatial contour should be $4L_t$. In Fig.~\ref{fig:OTOC}(b), we stretch the spatial contour into a circle, which is correspondingly marked by the same set of colors.
 
ii) Boundary operators: In Eq.~\eqref{Fphi}, $e^{\pm i \phi \hat{W}}$ is an operator that uniformly acts on all spatial sites. Then, after performing space-time duality, the dual operator again takes the form $e^{\pm i \phi \hat{w}}$, which acts as a time independent operator. In the double Keldysh contour, $e^{i\phi\hat{W}}$ and $e^{-i\phi\hat{W}}$ are separated by $2L_t$. Therefore, after space-time duality, $e^{\pm i \phi \hat{w}}$ act on two endpoints of a diameter in the spatial contour, which are denoted by squares in Fig.~\ref{fig:OTOC}. Therefore, these two operators are considered as the boundary operators in the dual picture. When $\phi=0$, $e^{i\phi\hat{W}}$ and $e^{\pm i\phi\hat{w}}$ are both identity operators. When we use Eq.~\eqref{Fphi} to diagnose quantum chaos, we concern the sensitivity of $\mathcal{F}$ when $\phi$ deviates from zero. In the dual picture, $e^{\pm i \phi \hat{w}}$ becomes the boundary operators, and this measures the sensitivity to boundary conditions, which is attributed to the non-Hermitian boundary effect.

iii) Equal time correlator: In Eq.~\eqref{Fphi}, $\hat{O}$ is a spatial local operator, and therefore, the space-time dual of Eq.~\eqref{Fphi} can be viewed as an equal time correlator of two bulk operators $\hat{\tilde{O}}$, where $\hat{\tilde{O}}$ is the space-time dual of $\hat{O}$. Two $\hat{\tilde{O}}$ operators are separated by $2L_t$ in the double Keldysh contour, and after space-time duality, they also sit at two endpoints of a diameter, as denoted by triangles in Fig.~\ref{fig:OTOC}. The spatial separation between the bulk operator $\hat{\tilde{O}}$ and the boundary operator ${e^{\pm i \phi \hat{w}}}$ is $L_t$. In quantum chaos, we concern the sensitivity to $\phi$ for long evolution steps $L_t$, therefore, in the dual picture, we concern the sensitivity for large spatial separation $L_t$ between the bulk operator and the boundary operator.  

The discussions above highlight the main feature of the space time duality of $\mathcal{F}(\phi)$. It can be shown more rigorously that the space-time duality of $\mathcal{F}(\phi)$ can be written as \cite{supple}\begin{equation} \label{Fphid}
\mathcal{F}(\phi)=\frac{\text{Tr}_{L_{\tilde{r}}}\left[(\hat{V}\hat{\mathcal{B}}(\phi))^{L_{\tilde{t}}}\hat{\tilde{O}}_{L_t+1}\hat{\tilde{O}}_{3L_t+1}\right] }{\text{Tr}_{L_{\tilde{r}}}\left[(\hat{V}\hat{\mathcal{B}}(\phi))^{L_{\tilde{t}}}\right]}.
\end{equation}
Here $\hat{\mathcal{B}}(\phi)=e^{i \phi \hat{w}_{2L_t+1}}e^{- i \phi \hat{w}_{1}}$ denotes the boundary operator in the dual picture. $\hat{V}$ in Eq.~\eqref{Fphid} is related to $\hat{U}$ in Eq.~\eqref{Fphi} via space-time duality. Hence, we have now mapped Eq.~\eqref{Fphi} into an equal-time correlator under a non-unitary evolution and in the presence of a boundary term. Nevertheless, we note that this correlator is not a standard two-point correlator in real time \cite{Zhang}. Quantum chaotic behavior in Eq.~\eqref{Fphi} is mapped to the sensitivity on the boundary parameter for large separation between bulk and boundary operators.     

{\color{blue}\emph{Numerical Results.}} Here we set $\hat{U}$ as given by Eq.~\eqref{bigU} and $\hat{V}$ behaves as Eq.~\eqref{bigV} \cite{supple}. Moreover, we choose $\hat{O}$ as $\hat{\sigma}_1^z$ and $\hat{W}$ as $\sum_{i=1}^{L_r}\hat{\sigma}^z_i$. The numerical results of $\mathcal{F}(\phi)$, as well as $\partial^2\mathcal{F}(\phi)/\partial\phi^2|_{\phi=0}$, are shown in Fig. \ref{fig:result}(a1),(b1) and (c1). We can see the sensitivity to $\phi$ even for large $L_t$. Here we would like to provide further evidence that this sensitivity of $\mathcal{F}(\phi)$ to $\phi$ can be interpreted as the non-hermitian boundary effect. To this end, we can artificially change the parameters $\tilde{J_x}$, $\tilde{J_z}$ and $h$ in $\hat{V}$ to be purely imaginary, such that $\hat{V}$ behaves as $e^{-\hat{H}}$, where $\hat{H}$ is a hermitian operator. Thus, the modified Eq.~\eqref{Fphid} can be viewed as the equal-time correlator of a statistical Hermitian system, and this modification eliminates the non-hermiticity in Eq.~\eqref{Fphid}. We plot this modified $\mathcal{F}(\phi)$, as well as $\partial^2\mathcal{F}(\phi)/\partial\phi^2|_{\phi=0}$, in Fig.~\ref{fig:result}(a2,b2,c2), and we should contrast Fig.~\ref{fig:result}(a2,b2,c2) with Fig.~\ref{fig:result}(a1,b1,c1).

i) In Fig.~\ref{fig:result}(a1), we plot $\mathcal{F}(\phi)$ as a function of $\phi$ for different $L_t$. One can see that, for large $L_t$, $\mathcal{F}(\phi)$ approaches $\cos(\phi)^{2L_r}$. This gives rise to OTOC as $2L_r$, which is consistent with the fully scrambled limit. In the fully scrambled limit, $\hat{\sigma}^z_1(L_t)$ uniformly populates the entire operator space, then $\text{Tr}_{L_r}\left[|[\hat{O}(L_t),\hat{W}]|^2\right]/2^{L_r}$ in Eq.~\eqref{derivative} approaches $2L_r$. In contrast, we show in Fig.~\ref{fig:result}(a2) that when $L_t$ is large enough, the modified $\mathcal{F}(\phi)$ approaches a constant independent of $\phi$. 

ii) In Fig.~\ref{fig:result}(b), we plot $\mathcal{F}(\phi)$ as a function of $L_t$ for different $\phi$. It is quite clear in this plot that, even for large $L_t$, $\mathcal{F}(\phi)$ also strongly depends on $\phi$. In contrast, Fig.~\ref{fig:result}(b2) shows that, for the modified $\mathcal{F}(\phi)$, the differences between $\mathcal{F}(\phi)$ with different $\phi$ become smaller when $L_t$ increases. 

iii) In Fig.~\ref{fig:result}(c1), we show the OTOC obtained from $\partial^2\mathcal{F}(\phi)/\partial\phi^2|_{\phi=0}$. The OTOC increases as $L_t$ increases, until it saturates to a finite non-zero value for large enough $L_t$, which is due to the finite size effect, and the saturation value is consistent with the fully scrambled limit of the finite Hilbert space. In contrast, Fig.~\ref{fig:result}(c2) shows that, for the modified $\mathcal{F}(\phi)$, the derivative $\partial^2\mathcal{F}(\phi)/\partial\phi^2|_{\phi=0}$ approaches zero as $L_t$ increases. 

All these results show that, when the non-hermiticity effect is mostly eliminated, the correlator of two bulk operators is no longer sensitive to the boundary parameter $\phi$ when the separation $L_t$ between the bulk operators and the boundary is large enough. This is consistent with our intuition of a hermitian system where the boundary effect should not significantly affect properties deeply in the bulk. In other words, it supports the claim the interpretation of the  results shown in Fig.~\ref{fig:result}(a1,b1,c1) are due to the non-hermiticity in the dual picture. 

{\color{blue}\emph{Discussions.}} In summary, we have established the connection between quantum chaos characterized by OTOC in hermitian quantum systems and the sensitivity to boundary conditions in non-hermitian systems. This study can stimulate many future research topics, and as examples, we would like to conclude this work by making the following two remarks. 

First, the non-hermitian skin effect has been mostly studied in non-interacting systems so far.  Here we note that the non-unitary evolution studied here in the dual picture cannot be viewed as free dynamics. Nevertheless, the sensitivity to boundary parameters still holds. This means that the sensitivity to boundary parameters is a generic feature of non-hermitian systems beyond single-particle physics. In other words, our study can also be viewed as an alternative route to generalize the skin effect to interacting non-hermitian systems. 

Secondly, here we have only considered the chaotic dynamics in hermitian systems. It is known what the OTOC behaves differently in non-chaotic systems, such as systems with many-body localization \cite{MBL1,MBL2,MBL3,MBL4,MBL5}. Therefore, it is natural to ask how the difference between chaotic and non-chaotic quantum system manifest itself in the dual non-unitary dynamics. This can shed new light on understanding boundary effect in non-hermitian system. 

\textit{Acknowledgements.} This work is supported by Beijing Outstanding Young Scientist Program and NSFC Grant No. 11734010.

\end{document}


\title{Supplementary material: Space-Time Duality between Quantum Chaos and Non-Hermitian Boundary Effect}
\author{Tian-Gang Zhou}
\thanks{They contribute equally to this work. }
\affiliation{Institute for Advanced Study, Tsinghua University, Beijing,100084, China}

\author{Yi-Neng Zhou}
\thanks{They contribute equally to this work. }
\affiliation{Institute for Advanced Study, Tsinghua University, Beijing,100084, China}

\author{Pengfei Zhang}
\thanks{PengfeiZhang.physics@gmail.com}
\affiliation{Institute for Quantum Information and Matter and Walter Burke Institute for Theoretical Physics, California Institute of Technology, Pasadena, California 91125, USA}

\author{Hui Zhai}
\thanks{hzhai@mail.tsinghua.edu.cn}
\affiliation{Institute for Advanced Study, Tsinghua University, Beijing,100084, China}
\date{\today}	
\maketitle

In this supplementary material, we present details of the space-time duality and the choice of the non-Hermitian and Hermitian models.

\section{Space-time duality of the two-qubit gate}
\label{sec:twoqubit}
First of all, the basic element of the space-time duality can be revealed on the two-qubit gate. As demonstrated in the Fig.~\ref{fig:ut_vr_brickwall}(a, b), the two-qubit gate evolving along the temporal direction is defined as:
		\begin{equation}
		\begin{aligned}
			\hat{u}=\exp\left(i  J_{z} \hat{\sigma}^z_r \hat{\sigma}^z_{r+1} \right) \exp \left(i J_{x} (\hat{\sigma}^x_r+\hat{\sigma}^x_{r+1}) \right) \exp \left(i h (\hat{\sigma}^z_r+\hat{\sigma}^z_{r+1})+i  J_{z} \hat{\sigma}^z_r \hat{\sigma}^z_{r+1} \right).
		\end{aligned}
	\end{equation}  
To apply the space-time duality, we evaluate the matrix elements of the two-qubit gate $\hat{u}$ in $\hat{\sigma}^z$ eigen-basis $ {|s_{r,t}\rangle}$. We calculate each part separately in the subsec.~\ref{ssec:sigmax},~\ref{ssec:sigmaz} and summarize the result in the subsec.~\ref{ssec:u}.
	
\subsection{Space-time dual of the $\hat{\sigma}^x_r$ term}
\label{ssec:sigmax}
For the $\exp(i J_{x} \hat{\sigma}^x_r)$ term, the matrix elements read as:
		\begin{equation}
		\begin{aligned}
			\langle s_{r,t+1}\vert \exp(i J_{x} \hat{\sigma}^x_r)\vert s_{r,t}\rangle \equiv k\exp(i \tilde{J_{z}} s_{r,t} s_{r,t+1} ),
		\end{aligned}
	\end{equation}  
where the classical variable $s_{r,t}$ can choose either $+1$ or $-1$ and
\begin{equation}\label{eq:tildeJz}
	k^2=\frac{i}{2}\sin(2J_x), \ \ \ \tilde{J}_{z} = -\frac{\pi}{4} + \frac{i}{2} \ln(\tan J_x).
\end{equation}
 
Inspired by the spirit of the space-time duality, we exchange the roles of the $t$ and $r$ with $\tilde{r}\equiv t,\ \tilde{t}\equiv r$. Then we introduce a new classical variable $\tilde{s}$. Specifically, we have 
\begin{equation}\label{eq:srt}
 s_{r,t}=s_{\tilde{t},\tilde{r}} \equiv \tilde{s}_{\tilde{r},\tilde{t}},
\end{equation}
where $\tilde{s}_{\tilde{r},\tilde{t}}$ is defined as swapping the first and second subscripts of $s_{\tilde{t},\tilde{r}}$.
Using this definition, the matrix elements can be equivalently expressed as
	\begin{equation}
	\begin{aligned}
		\langle s_{r,t+1}\vert \exp(i J_{x} \hat{\sigma}^x_r)\vert s_{r,t}\rangle &\equiv k\exp(i \tilde{J_{z}} s_{r,t} s_{r,t+1} )\\
		&=k\exp(i \tilde{J_{z}} \tilde{s}_{\tilde{r},\tilde{t}}  \tilde{s}_{\tilde{r}+1,\tilde{t}})\\
		&=k\langle \tilde{s}_{\tilde{r},\tilde{t}},\tilde{s}_{\tilde{r}+1,\tilde{t}}\vert \exp(i \tilde{J_{z}}  \hat{\sigma}^z_{\tilde{r} }\hat{\sigma}^z_{\tilde{r}+1})\vert \tilde{s}_{\tilde{r},\tilde{t}},\tilde{s}_{\tilde{r}+1,\tilde{t}}\rangle .
	\end{aligned}
\end{equation}  

\subsection{Space-time dual of the $\hat{\sigma}^z_r$ term}
\label{ssec:sigmaz}
Similarly, for the $\exp(i  J_{z} \hat{\sigma}^z_r \hat{\sigma}^z_{r+1}) $ term:
\begin{equation}
	\begin{aligned}
		\langle s_{r,t},s_{r+1,t}\vert \exp(i J_{z} \hat{\sigma}^z_r \hat{\sigma}^z_{r+1})\vert s_{r,t},s_{r+1,t}\rangle &=\exp(i J_{z} s_{r,t} s_{r+1,t} )\\
		&=\exp(i J_{z} \tilde{s}_{\tilde{r},\tilde{t}}  \tilde{s}_{\tilde{r},\tilde{t}+1})\\
		&\equiv \langle \tilde{s}_{\tilde{r},\tilde{t}+1}\vert k^{\prime} \exp(i \tilde{J_{x}}  \hat{\sigma}^x_{\tilde{r} })\vert \tilde{s}_{\tilde{r},\tilde{t}}\rangle ,
	\end{aligned}
\end{equation}  
with
\begin{equation}\label{eq:tildeJx}
	k'^2=\frac{2}{i\sin(2\tilde{J}_x)}, \ \ \ \tilde{J}_{x} = \arctan(-i e^{-2i J_z}).
\end{equation}

\begin{figure}[t]
	\centering
	\includegraphics[width=0.6\linewidth]{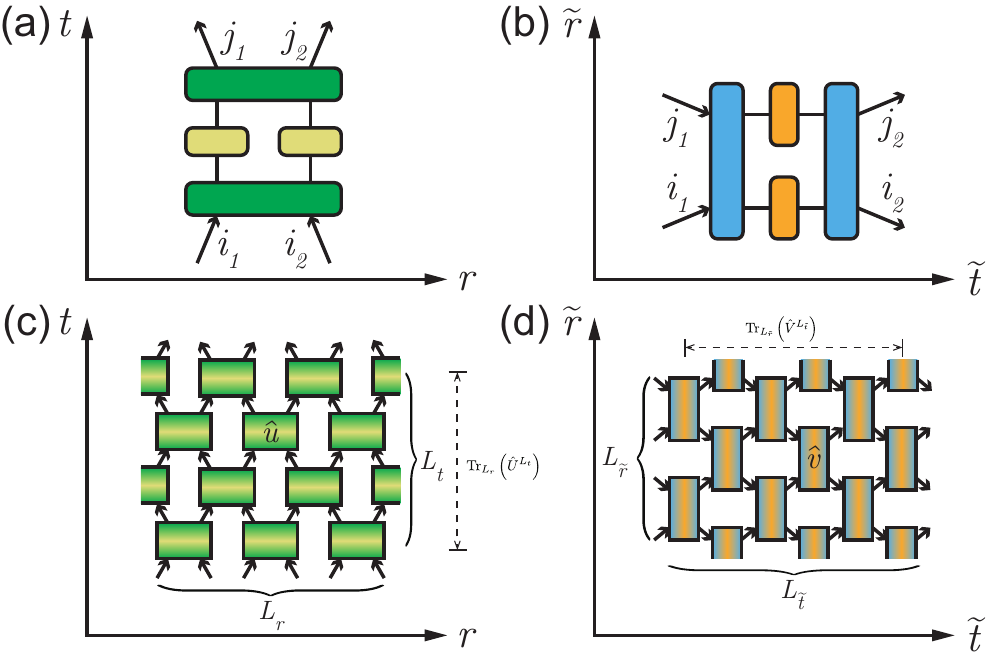}
	\caption{Schematic of space-time duality of quantum circuit. To be concrete, we consider more general case and the convention is slightly different with the main text. (a, b): Space-time duality between two-qubit quantum circuits $\hat{u}$ (left) and $\hat{v}$ (right) with $v_{i_1,j_1}^{i_2,j_2}=u_{i_1,i_2}^{j_1,j_2}$. Green and blue box respectively represent two-qubit gate $e^{iJ_z\hat{\sigma}^z_1\hat{\sigma}^z_2}$ and $e^{i\tilde{J}_z\hat{\sigma}^z_1\hat{\sigma}^z_2}$. Yellow and orange box respectively represent the combination of the single-qubit gates $e^{iJ_x\hat{\sigma}_x} e^{i h\hat{\sigma}_z}$ and $e^{i\tilde{J}_x\hat{\sigma}_x} e^{i h\hat{\sigma}_z}$, with the relation between $J_z$ and $\tilde{J}_z$, and relation between $J_x$ and $\tilde{J}_x$ given by Eq.~\eqref{eq:tildeJz} and \eqref{eq:tildeJx}. (c, d): Space-time duality between two general operators $\hat{U}$ and $\hat{V}$, with $L_{\tilde{r}}=L_t$ and $L_{\tilde{t}}=L_r$. Here we illustrate the $\hat{U}$ and $\hat{V}$ in the form of brick-walls, for the convenience of the prove in the sec.~\ref{sec:trace}.}
	\label{fig:ut_vr_brickwall}
\end{figure}

\subsection{Space-time dual of the $\hat{u}$}
\label{ssec:u}
Combining the $\hat{\sigma}^z_r$ term and the $\hat{\sigma}^x_r$ term together, we express the two-qubit gate $\hat{u}$ in the basis of the new classical variable $\tilde{s}$
\begin{equation}
	\begin{aligned}
		& \langle s_{r,t+1},s_{r+1,t+1}\vert \hat{u} \vert s_{r,t},s_{r+1,t}\rangle \\
		&=\langle s_{r,t+1},s_{r+1,t+1}\vert \exp(i  J_{z} \hat{\sigma}^z_r \hat{\sigma}^z_{r+1}) \exp(i J_{x} (\hat{\sigma}^x_r+\hat{\sigma}^x_{r+1})) \exp(i h (\hat{\sigma}^z_r+\hat{\sigma}^z_{r+1})+i  J_{z} \hat{\sigma}^z_r \hat{\sigma}^z_{r+1} )\vert s_{r,t},s_{r+1,t}\rangle \\
		&=\exp(i J_{z} s_{r,t+1}  s_{r+1,t+1})k\exp(i \tilde{J_{z}} s_{r,t}  s_{r,t+1})k\exp(i \tilde{J_{z}} s_{r+1,t}  s_{r+1,t+1}) \exp(i h (s_{r,t}+s_{r+1,t})+i J_{z} s_{r,t}  s_{r+1,t})\\
		&=\exp(i J_{z} \tilde{s}_{\tilde{r}+1,\tilde{t}}  \tilde{s}_{\tilde{r}+1,\tilde{t}+1})k\exp(i \tilde{J_{z}} \tilde{s}_{\tilde{r},\tilde{t}}  \tilde{s}_{\tilde{r}+1,\tilde{t}})k\exp(i \tilde{J_{z}} \tilde{s}_{\tilde{r},\tilde{t}+1}  \tilde{s}_{\tilde{r}+1,\tilde{t}+1})\exp(ih (\tilde{s}_{\tilde{r},\tilde{t}} +\tilde{s}_{\tilde{r},\tilde{t}+1} )  +i J_{z} \tilde{s}_{\tilde{r},\tilde{t}}  \tilde{s}_{\tilde{r},\tilde{t}+1})\\
		&=k^2\exp(i \tilde{J_{z}} \tilde{s}_{\tilde{r},\tilde{t}+1}  \tilde{s}_{\tilde{r}+1,\tilde{t}+1})\exp(i J_{z} \tilde{s}_{\tilde{r}+1,\tilde{t}}  \tilde{s}_{\tilde{r}+1,\tilde{t}+1})\exp(i J_{z} \tilde{s}_{\tilde{r},\tilde{t}}\tilde{s}_{\tilde{r},\tilde{t}+1}) \exp(ih (\tilde{s}_{\tilde{r},\tilde{t}} +\tilde{s}_{\tilde{r},\tilde{t}+1} ))\exp(i \tilde{J_{z}} \tilde{s}_{\tilde{r},\tilde{t}}  \tilde{s}_{\tilde{r}+1,\tilde{t}})\\
		&=k^2k'^2\langle \tilde{s}_{\tilde{r},\tilde{t}+1},\tilde{s}_{\tilde{r}+1,\tilde{t}+1}\vert  \hat{v} \vert \tilde{s}_{\tilde{r},\tilde{t}},\tilde{s}_{\tilde{r}+1,\tilde{t}}\rangle .
	\end{aligned}
\end{equation}  
Here, the fourth line is derived by rewriting $s$ as the new classical variable $\tilde{s}$, and the fifth line rearranges the terms in the row above. Followed by these steps, $\hat{v}$ is defined as
\begin{equation}
	\begin{aligned}
		\hat{v}=\exp(i  \tilde{J}_{z} \hat{\sigma}^z_{\tilde{r}} \hat{\sigma}^z_{\tilde{r}+1}) \exp(i \tilde{J}_{x} (\hat{\sigma}^x_{\tilde{r}+1}+\hat{\sigma}^x_{\tilde{r}})) \exp(i h (\hat{\sigma}^z_{\tilde{r}}+\hat{\sigma}^z_{\tilde{r}+1})+i  \tilde{J}_{z} \hat{\sigma}^z_{\tilde{r}} \hat{\sigma}^z_{\tilde{r}+1} ).
		\end{aligned}
\end{equation}  
Thus, we obtain the form of $\hat{v}$ which is dual to the two-qubit gate $\hat{u}$ considered at the beginning of this section.

\section{space-time duality of trace of unitary evolution} \label{sec:trace}

	We further derive the space-time duality of the trace of a unitary evolution. Illustrated in the Fig.~\ref{fig:ut_vr_brickwall}(c,d), we consider the 1-dimensional $L_r$ quantum circuit with periodic boundary condition(PBC), evolving along the temporal direction with $L_t$ steps. The time evolution of the quantum circuit reads as
\begin{equation}
	\hat{U}=
	\begin{cases}
				\exp(i  \sum_{r\in\text{odd}}^{L_r} J_{z} \hat{\sigma}^z_{r} \hat{\sigma}^z_{r+1}) \exp(i J_{x} \sum_{r=1}^{L_r}\hat{\sigma}^x_{r}) \exp(i h\sum_{r=1}^{L_r} \hat{\sigma}^z_{r}+i  J_{z}\sum_{r\in\text{odd}}^{L_r} \hat{\sigma}^z_{r} \hat{\sigma}^z_{r+1} ) & t=\text{odd} \\
				\exp(i  \sum_{r\in\text{even}}^{L_r} J_{z} \hat{\sigma}^z_{r} \hat{\sigma}^z_{r+1}) \exp(i J_{x} \sum_{r=1}^{L_r}\hat{\sigma}^x_{r}) \exp(i h\sum_{r=1}^{L_r} \hat{\sigma}^z_{r}+i  J_{z}\sum_{r\in\text{even}}^{L_r} \hat{\sigma}^z_{r} \hat{\sigma}^z_{r+1} ) & t=\text{even}. \\
	\end{cases}
\end{equation}	
We distinguish the even and odd sites here, since the nearest neighbor Ising interaction $\exp\left( i \sum_{r=1}^{L_r} \left(J_{z} \hat{\sigma}^z_r \hat{\sigma}^z_{r+1}\right) \right)$ is treated as two layers of two-qubit gates with brick-wall structure, depicted in the Fig.~\ref{fig:ut_vr_brickwall}(a,c). By applying the same technique in sec.~\ref{sec:twoqubit}, we insert $\hat{\sigma}^z$ eigen-basis $| s_{r,t} \rangle$ in the trace of the unitary evolution with $L_t$ steps, and then perform the space-time rotation
\begin{equation}
	\begin{aligned}
		\Tr_{L_r} \left( \hat{U}^{L_t} \right)
		&=\prod_{t\in\text{odd}}^{L_t} \exp(i J_{z} \sum_{r\in\text{odd}}^{L_r}s_{r,t+1}  s_{r+1,t+1})k^{L_r}\exp(i \sum_{r=1}^{L_r}\tilde{J_{z}} s_{r,t}  s_{r,t+1})\exp(i h \sum_{r=1}^{L_r}s_{r,t}) \exp(i J_{z} \sum_{r\in\text{odd}}^{L_r}s_{r,t}  s_{r+1,t})\\
				&\ \ \ \prod_{t\in\text{even}}^{L_t} \exp(i J_{z} \sum_{r\in\text{even}}^{L_r}s_{r,t+1}  s_{r+1,t+1})k^{L_r}\exp(i \sum_{r=1}^{L_r}\tilde{J_{z}} s_{r,t}  s_{r,t+1})\exp(i h \sum_{r=1}^{L_r}s_{r,t}) \exp(i J_{z} \sum_{r\in\text{even}}^{L_r}s_{r,t}  s_{r+1,t})\\
		&=k^{L_r L_t}\prod_{t=1}^{L_t} \prod_{r=1}^{L_r} \exp(i J_{z} s_{r,t+1}  s_{r+1,t+1})\exp(i \tilde{J_{z}} s_{r,t}  s_{r,t+1})\exp(i h s_{r,t}) \\
		&=k^{L_{\tilde{t}} L_{\tilde{r}}} \prod_{\tilde{t}=1}^{L_{\tilde{t}}} \prod_{\tilde{r}=1}^{L_{\tilde{r}}} \exp(i J_{z} \tilde{s}_{\tilde{r}+1,\tilde{t}}  \tilde{s}_{\tilde{r}+1,\tilde{t}+1}) \exp(i \tilde{J_{z}} \tilde{s}_{\tilde{r},\tilde{t}}  \tilde{s}_{\tilde{r}+1,\tilde{t}})\exp(ih \tilde{s}_{\tilde{r},\tilde{t}} )  \\
		&=k^{L_{\tilde{t}} L_{\tilde{r}}} \prod_{\tilde{t}\in \text{odd}}^{L_{\tilde{t}}}\exp(i J_{z} \sum_{\tilde{r}\in\text{odd}}^{L_{\tilde{r}}}\tilde{s}_{\tilde{r}+1,\tilde{t}}  \tilde{s}_{\tilde{r}+1,\tilde{t}+1}) \exp(i \tilde{J_{z}}\sum_{\tilde{r}=1}^{L_{\tilde{r}}} \tilde{s}_{\tilde{r},\tilde{t}}  \tilde{s}_{\tilde{r}+1,\tilde{t}})\exp(ih\sum_{\tilde{r}=1}^{L_{\tilde{r}}} \tilde{s}_{\tilde{r},\tilde{t}} ) \exp(i J_{z} \sum_{\tilde{r}\in\text{odd}}^{L_{\tilde{r}}}\tilde{s}_{\tilde{r},\tilde{t}}  \tilde{s}_{\tilde{r},\tilde{t}+1}) \\
			&\ \ \ \ k^{L_{\tilde{t}} L_{\tilde{r}}} \prod_{\tilde{t}\in \text{even}}^{L_{\tilde{t}}}\exp(i J_{z} \sum_{\tilde{r}\in\text{even}}^{L_{\tilde{r}}}\tilde{s}_{\tilde{r}+1,\tilde{t}}  \tilde{s}_{\tilde{r}+1,\tilde{t}+1}) \exp(i \tilde{J_{z}}\sum_{\tilde{r}=1}^{L_{\tilde{r}}} \tilde{s}_{\tilde{r},\tilde{t}}  \tilde{s}_{\tilde{r}+1,\tilde{t}})\exp(ih\sum_{\tilde{r}=1}^{L_{\tilde{r}}} \tilde{s}_{\tilde{r},\tilde{t}} ) \exp(i J_{z} \sum_{\tilde{r}\in\text{even}}^{L_{\tilde{r}}}\tilde{s}_{\tilde{r},\tilde{t}}  \tilde{s}_{\tilde{r},\tilde{t}+1}) \\
		&=	\Tr_{L_{\tilde{r}}} \left( \hat{V}^{L_{\tilde{t}}}  \right)\\
	\end{aligned}
\end{equation}
with $\hat{V}$ defined as
	\begin{equation}
		\hat{V}=
		\begin{cases}
			(k k^{\prime})^{L_{\tilde{r}}}\exp(i  \sum_{\tilde{r}\in\text{odd}}^{L_{\tilde{r}}}\tilde{J}_{z} \hat{\sigma}^z_{\tilde{r}} \hat{\sigma}^z_{\tilde{r}+1}) \exp(i \tilde{J}_{x} \sum_{\tilde{r}=1}^{L_{\tilde{r}}}\hat{\sigma}^x_{\tilde{r}}) \exp(i h\sum_{\tilde{r}=1}^{L_{\tilde{r}}} \hat{\sigma}^z_{\tilde{r}}+i  \tilde{J}_{z}\sum_{\tilde{r}\in\text{odd}}^{L_{\tilde{r}}} \hat{\sigma}^z_{\tilde{r}} \hat{\sigma}^z_{\tilde{r}+1} ) & t=\text{odd}. \\
						(k k^{\prime})^{L_{\tilde{r}}}\exp(i  \sum_{\tilde{r}\in\text{even}}^{L_{\tilde{r}}}\tilde{J}_{z} \hat{\sigma}^z_{\tilde{r}} \hat{\sigma}^z_{\tilde{r}+1}) \exp(i \tilde{J}_{x} \sum_{\tilde{r}=1}^{L_{\tilde{r}}}\hat{\sigma}^x_{\tilde{r}}) \exp(i h\sum_{\tilde{r}=1}^{L_{\tilde{r}}} \hat{\sigma}^z_{\tilde{r}}+i  \tilde{J}_{z}\sum_{\tilde{r}\in\text{even}}^{L_{\tilde{r}}} \hat{\sigma}^z_{\tilde{r}} \hat{\sigma}^z_{\tilde{r}+1} ) & t=\text{even}. \\
		\end{cases}
	\end{equation} 
where $L_t=L_{\tilde{r}},  \ \ L_r=L_{\tilde{t}}$. Additionally, we notice the significance of the PBC, since the spatially periodic boundary condition is exactly matched with the trace operation along the temporal direction after space-time duality.

\section{space-time duality of correlation function $\mathcal{F}(\phi)$}
\begin{figure}[t]
	\centering
	\includegraphics[width=0.35\linewidth]{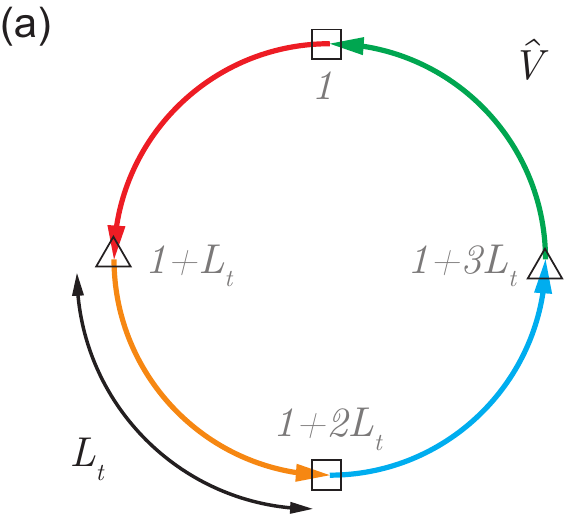}
	\caption{The double Keldysh contour demonstrated in the Fig.~2(b) of the main text but with numbering. Squares and triangles label $e^{\pm i\phi\hat{W}}$ and $\hat{O}$ acting on different spatial points of the spatial contour in the dual circuit respectively. The distance between triangle and square is $L_t$. Boundary terms $e^{\pm i\phi\hat{W}}$ are labeled by $1$ and $2L_t+1$ separately, whereas two local operators $\hat{O}$ are labeled by $L_t+1$ and $3L_t+1$.} 
	\label{fig:otoc_number}
\end{figure}

\subsection{Non-Hermitian system the space-time dual}
We investigate the space-time duality of the correlation function $\mathcal{F}(\phi)$, whose second-order derivative of $\phi$ is the out-of-time-ordered commutator(OTOC). $\mathcal{F}(\phi)$ is defined as
\begin{equation}
	\mathcal{F}(\phi)=\frac{1}{2^{L_r}}\text{Tr}_{L_r}\left[\hat{O}(L_t)e^{i\phi\hat{W}}\hat{O}^{\dag}(L_t)e^{-i\phi\hat{W}}\right].
\end{equation}
In our numerical result, we choose $\hat{W}=\sum_r \hat{\sigma}_r^z$ as an operator that uniformly acts on all spatial sites, and $\hat{O}= \hat{\sigma}_1^z$ which is a spatial local operator. Additionally, after writing the operator in the Heisenberg representation explicitly, we have
\begin{equation}
	\mathcal{F}(\phi)=\frac{1}{2^{L_r}}\text{Tr}_{L_r}\left[(\hat{U})^{\dag L_t}\hat{\sigma}_1^z(\hat{U})^{L_t}e^{i\phi\sum_r \hat{\sigma}_r^z}(\hat{U})^{\dag L_t}\hat{\sigma}_1^z(\hat{U})^{L_t}e^{-i\phi\sum_r \hat{\sigma}_r^z}\right]. \label{Fphi}
\end{equation}
Following the same procedure in sec.~\ref{sec:twoqubit}, we expand the $\mathcal{F}(\phi)$ in the diagonal basis $| {s_{r,t}} \rangle$
\begin{equation}
	\begin{aligned}
		\mathcal{F}(\phi)&=\frac{1}{2^{L_r}}k^{4L_r L_t}s_{L_t+1,1}  s_{3L_t+1,1}\prod_{r=1}^{L_r} \left[\exp(i  \sum_{t=1}^{4L_t}\mathcal{J}_{z,t}s_{r,t}  s_{r+1,t})\exp(i \sum_{t=1}^{4L_t} \tilde{\mathcal{J}}_{z,t}s_{r,t}  s_{r,t+1})\exp(i \sum_{t=1}^{4L_t}\tilde{h}_{t} s_{r,t})e^{-i\phi s_{r,1}}e^{i\phi s_{r,2L_t+1}}\right].\\
	\end{aligned}
\end{equation}
The labeling is according to the convention illustrated in the Fig.~\ref{fig:otoc_number}. Furthermore, $\mathcal{F}(\phi)$ can be represented in terms of the new classical variable $\tilde{s}$: 
\begin{equation}
	\label{eq:Fotocclassical}
	\begin{aligned}
		\mathcal{F}(\phi)&=\frac{1}{2^{L_{\tilde{t}}}} k^{L_{\tilde{t}} L_{\tilde{r}}}\tilde{s}_{1,L_t+1}  \tilde{s}_{1,3L_t+1}\prod_{\tilde{t}=1}^{L_{\tilde{t}}} \left[ \exp(i  \sum_{\tilde{r}=1}^{L_{\tilde{r}}} \mathcal{J}_{z,\tilde{r}} \tilde{s}_{\tilde{r},\tilde{t}}  \tilde{s}_{\tilde{r},\tilde{t}+1}) \exp(i \sum_{\tilde{r}=1}^{L_{\tilde{r}}} \tilde{\mathcal{J}}_{z,\tilde{r}}\tilde{s}_{\tilde{r},\tilde{t}}  \tilde{s}_{\tilde{r}+1,\tilde{t}})\exp(i  \sum_{\tilde{r}=1}^{L_{\tilde{r}}}\tilde{h}_{\tilde{r}} \tilde{s}_{\tilde{r},\tilde{t}} )e^{-i\phi \tilde{s}_{1,\tilde{t}}}e^{i\phi \tilde{s}_{2L_t+1,\tilde{t}}}\right],\\
	\end{aligned}
\end{equation}
where $4L_t=L_{\tilde{r}},\ L_r=L_{\tilde{t}}$, and the parameters are shown in the Eq.~\eqref{eq:type1}. Transforming it into the trace formula, we have
\begin{equation}
	\mathcal{F}(\phi)= \text{Tr}_{L_{\tilde{r}}}\left[(\hat{V}_{\RN{1}} \hat{\mathcal{B}}(\phi))^{L_{\tilde{t}}}\hat{\tilde{O}}_{L_t+1}\hat{\tilde{O}}_{3L_t+1}\right],
\end{equation}
with
\begin{equation}
	\label{eq:V1}
	\begin{aligned}
		\hat{V}_{\RN{1}} =& \frac{1}{2^{L_{\tilde{t}}}} (k k^{\prime})^{L_{\tilde{t}} L_{\tilde{r}}}
		\left[\prod_{j=0}^{3} \left(\hat{\sigma}^0_{1+j L_t}+\hat{\sigma}^x_{1+j L_t} \right) \right]
		\exp \left(\iu  \sum_{\tilde{r}=1}^{L_{\tilde{r}}} \tilde{\mathcal{J}}_{x,\tilde{r}} \hat{\sigma}^x_{\tilde{r}} \right) \exp \left(
		\iu \sum_{\tilde{r}=1}^{L_{\tilde{r}}} \tilde{\mathcal{J}}_{z,\tilde{r}} \hat{\sigma}^z_{\tilde{r}} \hat{\sigma}^z_{\tilde{r}+1} +
		\iu \sum_{\tilde{r}=1}^{L_{\tilde{r}}} \tilde{h}_{\tilde{r}} \hat{\sigma}^z_{\tilde{r}} \right),
	\end{aligned}
\end{equation}
and
\begin{equation}
	\label{eq:boundary}
	\hat{\mathcal{B}}(\phi)=e^{i \phi \hat{\sigma}_{2L_t+1}^z} e^{-i \phi \hat{\sigma}_{1}^z}.
\end{equation}
The parameters on the contour are
\begin{equation}
	\label{eq:type1}
	\def\arraystretch{1.0}
	\begin{tabular}{c|cccc}
		\hline
		&$\mathcal{J}_{z,\tilde{r}}$&$\tilde{\mathcal{J}}_{x,\tilde{r}}$ & $ \tilde{\mathcal{J}}_{z,\tilde{r}}$ & $ \tilde{h}_{\tilde{r}}$ \\ \hline
		$\tilde{r}=1$     &   $0$                                  &           $0$                                    & $ \tilde{J}_z$                                   &   $ 0$       \\ \hline
		$1<\tilde{r}<L_t+1$   &$J_z$                         &       $ \tilde{J}_x $                       &        $ \tilde{J}_z$                          &      $ h $       \\ \hline
		$\tilde{r}=L_t+1$    &  $0$                            &        $ 0 $                                      &             $ \tilde{J}'_z$             &       $ 0$       \\ \hline
		$L_t+1<\tilde{r}<2L_t+1$  &  $-J_z$            &    $ \tilde{J}'_x $                         &        $ \tilde{J}'_z$                     &      $ -h $      \\ \hline
		$\tilde{r}=2L_t+1$    &  $0$                          &       $ 0 $                                     &             $ \tilde{J}_z$                &       $ 0$       \\ \hline
		$2L_t+1<\tilde{r}<3L_t+1$ & $J_z$             &     $ \tilde{J}_x $                         &        $ \tilde{J}_z$                     &      $ h $      \\ \hline
		$\tilde{r}=3L_t+1$    &   $0$                        &       $ 0 $                                      &             $ \tilde{J}'_z$             &       $ 0$       \\ \hline
		$3L_t+1<\tilde{r}<4L_t+1$ & $-J_z$         &  $ \tilde{J}'_x $                            &       $ \tilde{J}'_z$                     &      $ -h $      \\ \hline
	\end{tabular}
	\text{\ \ \ with\ \ \ }
	\def\arraystretch{1.0}
	\begin{tabular}{cc}
		\hline
		&$\tilde{J}_{x} = \arctan(-\iu \exp(-2\iu J_z))$  \\ \hline
		&$\tilde{J}'_{x} = \arctan(-\iu \exp(2\iu J_z)) =  \tilde{J}_{x} + \pi/2$  \\  \hline
		&$\tilde{J}_{z} = -\pi/4 + \frac{i}{2} \ln(\tan J_x)$ \\ \hline
		&$\tilde{J}'_{z} = -\pi/4 + \frac{i}{2} \ln(\tan (-J_x)) =  \tilde{J}_{z} - \pi/2$ \\ \hline
	\end{tabular}
\end{equation}
From the Fig.~\ref{fig:otoc_number} and Eq.~\eqref{eq:type1}, we notice that the parameters of the qubits at $\tilde{r}=1,L_t+1,2L_t+1,3L_t+1$ are special, since these sites connect the forwards and backwards evolutions. We call these qubits as \textit{edge qubits} for later convenience. The Eq.~\eqref{eq:V1} shows, projectors $(\hat{\sigma}^0_{\tilde{t}}+\hat{\sigma}^x_{\tilde{t}})$ exist at each edge qubit, since the $\mathcal{J}_{z,\tilde{r}} = 0$ at these sites. We will give a proof in the sec.~\ref{sec:proj}.

Finally, we note that $\hat{V}$ in general describe a non-unitary evolution whereas $\hat{U}$ is unitary. Firstly, parameters $\{\tilde{J}_{x}, \tilde{J}_{z}, \tilde{J}'_{x}, \tilde{J}'_{z} \}$ in $\hat{V}$ have non-vanishing imaginary part, but parameters in $\hat{U}$ are all real number. Secondly, the projectors in Eq.~\eqref{eq:V1} are intrinsically non-unitary.

\subsection{Hermitian system}
To provide further evidence on the non-Hermitian boundary effect, we artificially change the parameters $\tilde{J_x}$, $\tilde{J_z}$ and $\tilde{h}$ in $\hat{V}$ to be purely imaginary and throw away the projectors at edge qubits, such that $\hat{V}$ behaves as $e^{-\hat{H}}$, where $\hat{H}$ is a Hermitian operator. Explicitly, the correlator $\mathcal{F}(\phi)$ can be formulated as
\begin{equation}
	\mathcal{F}(\phi)= \text{Tr}_{L_{\tilde{r}}}\left[(\hat{V}_{\RN{2}} \hat{\mathcal{B}}(\phi))^{L_{\tilde{t}}}\hat{\tilde{O}}_{L_t+1}\hat{\tilde{O}}_{3L_t+1}\right],
\end{equation}
\begin{equation}
	\label{eq:V2}
	\begin{aligned}
		\hat{V}_{\RN{2}} =\exp \left(-\beta  \sum_{\tilde{r}=1}^{L_{\tilde{r}}} \tilde{\mathcal{J}}_{x,\tilde{t}} \hat{\sigma}^x_{\tilde{r}} \right) \exp \left(
		-\beta \sum_{\tilde{r}=1}^{L_{\tilde{r}}} \tilde{\mathcal{J}}_{z,\tilde{t}} \hat{\sigma}^z_{\tilde{r}} \hat{\sigma}^z_{\tilde{r}+1} -\beta \sum_{\tilde{r}=1}^{L_{\tilde{r}}} \tilde{h}_{\tilde{t}} \hat{\sigma}^z_{\tilde{r}} \right),
	\end{aligned}
\end{equation}
where $\beta$ is the inverse temperature and $\hat{\mathcal{B}}(\phi)$ is defined in Eq.~\eqref{eq:V1}. We summarize the parameters on the contour in the following table
\begin{equation}
	\def\arraystretch{1.0}
	\begin{tabular}{c|ccc}
		\hline
		& $\tilde{\mathcal{J}}_{x,\tilde{r}}$ & $ \tilde{\mathcal{J}}_{z,\tilde{r}}$ & $ \tilde{h}_{\tilde{r}}$ \\ \hline
		$\tilde{r}=1$     &            $ J_x $            &             $ J_z$             &      $ h$       \\ \hline
		$1<\tilde{r}<L_t+1$   &       $ J_x $       &        $J_z $        &      $ h $       \\ \hline
		$\tilde{r}=L_t+1$    &            $ J_x $            &             $ J_z$             &       $ h$       \\ \hline
		$L_t+1<\tilde{r}<2L_t+1$  &      $ J_x $       &        $ J_z$        &      $ h $      \\ \hline
		$\tilde{r}=2L_t+1$    &            $ J_x $            &             $ J_z$             &       $ h$       \\ \hline
		$2L_t+1<\tilde{r}<3L_t+1$ &       $ J_x $       &        $ J_z$        &      $ h $      \\ \hline
		$\tilde{r}=3L_t+1$    &            $ J_x $            &             $ J_z$             &       $ h$       \\ \hline
		$3L_t+1<\tilde{r}<4L_t+1$ &      $ J_x $       &       $ J_z$        &      $ h $      \\ \hline
	\end{tabular}
\end{equation}
In our numerical calculation, we set $J_z=J_x=1,h=0.5, \beta=0.1$. In the high-temperature limit, the Eq.~\eqref{eq:V1} approximately acts as the thermal density matrix with Hermitian Hamiltonian $\hat{H}$. 

\section{Force projective measurement on edge qubits}
\label{sec:proj}

We discuss the details of the force projective measurement in the Eq.~\eqref{eq:V1}. First of all, we prove a proposition on a minimal case.

\textbf{Proposition}
	When $J_z = 0$, the two-qubit gate $\exp(\iu J_{z} \hat{\sigma}^z_{r} \hat{\sigma}^z_{r+1}) $ on the spatial sites $[r,r+1]$ is mapped to a single-qubit gate $(\hat{\sigma}^0+\hat{\sigma}^z)_{t}$ on the temporal site $t$.

\begin{proof}
	We consider a general two-qubit gate $\hat{u}_{i_1,i_2}^{o_1,o_2}$, where $i_1,i_2$ denote two input qubits and $o_1,o_2$ denote two output qubits. Via space-time duality, this gate is dual to another two-qubit gate, with
	\begin{equation}
		\hat{\tilde{u}}_{i_1,i_2}^{o_1,o_2} \equiv \hat{u}_{i_1,o_1}^{i_2,o_2}.
	\end{equation}
	Particularly, for the two-qubit gate $\exp\left( \iu J_{z} \hat{\sigma}^z_{r} \hat{\sigma}^z_{r+1} \right) $, we compute the matrix element in the $\hat{\sigma}^z$ eigen-basis  $\left(\vert-1,-1\rangle,\vert-1,1\rangle,\vert1,-1\rangle,\vert1,1\rangle \right)$
	\begin{equation}
		\hat{u}= \begin{pmatrix}
			\exp(\iu J_{z})& 0 & 0 &0\\
			0 &\exp(-\iu J_{z})& 0 & 0\\
			0 & 0&\exp(-\iu J_{z}) & 0\\
			0 & 0 & 0 &\exp(\iu J_{z})\\
		\end{pmatrix}.
	\end{equation}
	After space-time duality, we get
	\begin{equation}
			\hat{\tilde{u}} = \begin{pmatrix}
				\exp(\iu J_{z})& 0 & 0 &\exp(-\iu J_{z})\\
				0 &0 & 0 & 0\\
				0 & 0&0 & 0\\
				\exp(-\iu J_{z}) & 0 & 0 &\exp(\iu J_{z})\\
			\end{pmatrix}.
	\end{equation}
	Although $\tilde{u}$ is originally defined as a 4-dimensional matrix, it is easy to find that $\tilde{u}$ has a 2-dimensional irreducible representation whose basis are $\vert -1,-1\rangle$ and $\vert 1,1\rangle$. Thus, this two-qubit gate $\tilde u$ can be regarded as a single-qubit gate on time $t$, in the reduced basis $(\vert -1,-1\rangle,\ \vert 1, 1\rangle)$ with
	\begin{equation}
		\hat{\tilde{u}}=\begin{pmatrix}
			\exp(\iu J_{z}) &\exp(-\iu J_{z})\\
			\exp(-\iu J_{z}) &\exp(\iu J_{z})\\
		\end{pmatrix}.
	\end{equation}
	In the case of $J_z=0$, 
	\begin{equation}
		\hat{\tilde{u}}\Big|_{J_z=0}=
		\begin{pmatrix}
			1 &1\\
			1 &1\\
		\end{pmatrix}
		=\hat{\sigma}^0+\hat{\sigma}^x.
	\end{equation}
	This completes the proof of the proposition. 
\end{proof}
	
	Secondly, at four edge qubits, the backwards evolution $\hat{U}$ is adjacent to the forwards evolution $\hat{U}^{\dagger}$. It can be verified that these four cases 
	\begin{itemize}
		\item $(\hat{U}^{\dag})\hat{\sigma}_1^z(\hat{U})$
		\item $(\hat{U}) e^{i\phi\sum_r \hat{\sigma}_r^z}(\hat{U})^{\dagger}$
		\item $(\hat{U}^{\dag})\hat{\sigma}_1^z(\hat{U})$
		\item $(\hat{U}) e^{-i\phi\sum_r \hat{\sigma}_r^z}(\hat{U})^{\dagger}$
	\end{itemize}
	all lead to $\mathcal{J}_{z,\tilde{r}} = 0$ in the Eq.~\eqref{eq:Fotocclassical} effectively. Applying the theorem above, one can immediately obtain the force projective measurement $\left[\prod_{j=0}^{3} \left(\hat{\sigma}^0_{1+j L_t}+\hat{\sigma}^x_{1+j L_t} \right) \right]$ in the Eq.~\eqref{eq:V1}.